\documentclass[aip,reprint,longbibliography,superscriptaddress]{revtex4-1}

\usepackage{graphicx}
\usepackage{longtable}
\usepackage{array}
\usepackage{dcolumn}
\usepackage{amsmath}
\usepackage{amsfonts}
\usepackage{amssymb}
\usepackage{soul}
\usepackage{bm}
\usepackage{hyperref}

\allowdisplaybreaks[1]

\usepackage{color}

\begin{document}

\title{Broken symmetry $G_0W_0$ approach for the evaluation of exchange coupling constants}
\author{Akseli Mansikkam\"aki}
\email{akseli.mansikkamaki@oulu.fi}
\affiliation{NMR Research Unit, University of Oulu, P.\,O. Box 3000, FI-90014 Oulu, Finland.}

\author{Zhishuo Huang}
\affiliation{Theory of Nanomaterials Group, KU Leuven, Celestijnenlaan 200F, 3001 Leuven, Belgium}

\author{Naoya Iwahara}
\affiliation{Department of Chemistry, National University of Singapore, 3 Science Drive 3, 117543 Singapore}

\author{Liviu F. Chibotaru}
\email{liviu.chibotaru@kuleuven.be}
\affiliation{Theory of Nanomaterials Group, KU Leuven, Celestijnenlaan 200F, 3001 Leuven, Belgium}

\begin{abstract}
The applicability of a broken symmetry version of the $G_0W_0$ approximation to the calculation of isotropic exchange coupling constants has been studied. Using a simple H--He--H model system the results show a significant and consistent improvement of the results over both broken symmetry Hartree--Fock and broken symmetry density functional theory. In the case of more realistic bimetallic Cu(II) complexes, inclusion of the $G_0W_0$ correction does not lead to obvious improvement in the results. The discrepancies are explained by improved description of the interactions within the magnetic orbital space upon inclusion of the $G_0W_0$ corrections but deterioration of the description of charge- and spin-polarization effects outside the magnetic orbital space. Overall the results show that computational methods based on the $GW$ method have a potential to improve computational estimates of exchange coupling constants.
\end{abstract}

\maketitle

\section{Introduction}
The macroscopic magnetic properties of correlated insulators and paramagnetic molecules are determined by the nature of the local electronic states of individual paramagnetic ions and the magnetic interaction between these magnetic sites. The electronic structures of such materials are described by effective spin Hamiltonians constructed in terms of effective parameters describing the inter- and intrasite interactions.\cite{winter_2019a,khomskii_2014,whangbo_2013a,anderson_1963b} One of the most fundamental parameters describing the magnetic interaction between two sites is the isotropic exchange coupling constant $J_{ij}$, which describes the strength and type (ferromagnetic vs. anti-ferromagnetic) of the magnetic interaction in the Heisenberg--Dirac--van Vleck term\cite{heisenberg_1928a,dirac_1929a,van-vleck_1945a,slater_1953a,anderson_1963b} of the spin Hamiltonian
\begin{equation}
  \hat H_\mathrm{HDvV} = -\sum_{ij} J_{ij} \hat{\mathbf{S}}_i\cdot\hat{\mathbf{S}}_j\text{,}
\end{equation}
where $i$ and $j$ index the magnetic sites and $\hat{\mathbf{S}}_i$ is an effective local spin operator acting on site $i$.

The most widely used approach for the first-principles evaluations of exchange couplings is the density functional theory (DFT) within the broken symmetry (BS) formalism.\cite{noodleman_1981a,illas_2006a} The BS-DFT approach provides a convenient numerical recipe for the extraction of the exchange coupling parameters at reasonable computational costs. The method can be routinely applied to systems with hundreds of atoms and several magnetic sites. Despite its numerical success, the BS-DFT methodology remains problematic both in terms of its theoretical formulation and its practical application. No general consensus exists on how the BS state should be interpreted in the framework of Kohn--Sham (KS) DFT.\cite{illas_2006a,ruiz_2005a,illas_2006c,ruiz_2006a} Furthermore, it has been repeatedly shown that the results depend strongly on the choice of the approximation to the exchange--correlation (XC) functional.\cite{illas_2008a,illas_2008b,peralta_2010a,peralta_2011a,peralta_2012a} Whereas the sign of the exchange coupling constant is usually correctly produced, its magnitude can vary by several hundreds of percents depending on the choice of the XC approximation. These issues can be avoided by using high-level electron correlation methods based on configuration expansions of the full wave function.\cite{pantasiz_2018a,guihery_2014a} Although such methods are superior in accuracy and the results are straightforward to interpret, their application is severely limited by the computational costs which rise factorially with respect to the number of unpaired electrons.

The $GW$ approximation to many-body perturbation theory has been widely used in the field of solid-state physics to successfully describe various properties of correlated materials.\cite{hedin_1965a,gunnarsson_1998a,martin_2016} In recent years it has also been increasingly applied to molecular systems.\cite{van-setten_2013a,blase_2014a} Already at its simplest level of approximation, the so-called $G_0W_0$ approximation provides improved results on predictions of quantities such as molecular excitation spectra. The conceptual advantage of the $GW$ approach over DFT is that it does not depend on unknown quantities such as the XC functional. The $G_0W_0$ approach depends on the KS orbitals and the results still have an explicit dependence on the XC functional, but higher-level approximations to the $GW$ method, are completely independent of any preceding KS calculation. The $GW$ approach offers, both in principle and in practice, a way to extract values of $J_{ij}$ in a systematically improved manner without any reference to unknown mathematical entities.

In the present work we study the applicability of the $G_0W_0$ approximation to the evaluation of exchange coupling constants within the BS formalism. As a testing ground we employ the hypothetical H--He--H molecule, which has been widely used in the study of the BS-DFT methodology.\cite{yamaguchi_2000a,ruiz_1999a,ruiz_2005a,illas_2008a,illas_2008b} This system is complicated enough to include the most important physical contributions to the microscopic exchange coupling mechanism while still being simple enough to be treated exactly up to the basis set limit using high-level electron correlation methods. The system makes it possible to study the effects of the $G_0W_0$ approximation to the exchange coupling independent of other possible electronic structure features which might complicate the $G_0W_0$ calculation and introduce other sources of error. The length of the H--He bonds was varied from $1.5\,\text{\AA}$ to $2.5\,\text{\AA}$ in $0.1\,\text{\AA}$ steps. The short bond length limit represent strong exchange coupling comparable to covalent bonding whereas the long bond length limit represents vanishingly small exchange coupling. To test whether the conclusions based on results obtained for the H--He--H system are generalizable to more realistic systems, we also conduct calculations on simple experimentally characterized coordination complexes with two exchange-coupled Cu(II) ions.

\section{Computational methods}
The DFT and $G_0W_0$ calculations were carried out using the \textsc{Molgw} code version 2.A\cite{bruneval_2016a} specifically designed for molecular $GW$ calculations. The code was locally modified to allow a BS guess. First, a standard restricted DFT calculation was carried out and then the highest occupied and lowest unoccupied $\alpha$ and $\beta$ orbitals were mixed by $45^\circ$ and $-45^\circ$ rotations, respectively. This state was then converged to the BS state. The nature of the BS states were confirmed by examination of the molecular orbital coefficients and Mulliken atomic spin populations. The calculations were carried out using Hartree--Fock (HF) orbitals and KS orbitals obtained using various different exchange--correlation (XC) functionals as a starting point for the $G_0W_0$ calculation. The XC functionals included the pure generalized gradient approximation (GGA) BLYP\cite{becke_1988a,parr_1988a}, and the hybrid functionals B3LYP\cite{becke_1988a,parr_1988a,becke_1993a,stephens_1994a} and BHHLYP\cite{becke_1988a,parr_1988a,becke_1993b}. The two hybrid functionals contain $20\%$ and $50\%$ of exact exchange, respectively. Dunning's correlation consistent double-$\zeta$ quality basis cc-pVDZ\cite{dunning_1989a} was used in all calculations. A pseudopotential was used to treat the core electrons of the Cu(II) ions along with a corresponding valence basis set\cite{peterson_2005a}. The \textsc{Molgw} code only includes first-order algorithms for the convergence of the self-consistent field, which lead to problems in the case of more complicated electronic structures. Due to convergence issues HF reference was not calculated for the Cu(II) systems.

Reference values for the exchange coupling constants of the H--He--H system were evaluated using the complete active space self-consistent field (CASSCF) method\cite{roos_1987a,roos_2016} with all 15 orbitals and 4 electrons included in the active space. The calculation corresponds to an orbital-optimized full configuration interaction calculation and represents the most accurate possible solution to the Schr\"odinger equation within the basis set limit. The CASSCF calculations utilized the cc-pVDZ basis sets and were carried out using the \text{Orca} quantum chemistry software version 4.2.0\cite{neese_2017a}.

The geometries of the coordination complexes were extracted from the respective experimentally determined crystal structures and were used without further optimization. The structures are referred to by their reference codes in the \textsc{Cambridge Structure Database}.

The total $G_0W_0$ energies necessary for the extraction of the exchange coupling constants were evaluated using the Galitskii--Migdal formula\cite{galitskii_1958a} as implemented in \textsc{Molgw}. The exchange couplings were calculated using the Yamaguchi formula\cite{yamaguchi_1986a,yamaguchi_2000a}
\begin{equation}
  \label{E:projection}
  J = \frac{2(E_\mathrm{BS} - E_\mathrm{T})}{\langle {S_\mathrm{T}}^2\rangle - \langle {S_\mathrm{BS}}^2\rangle}\text{,}
\end{equation}
where $E_\mathrm{BS}$ and $E_\mathrm{T}$ are the energies of the BS and triplet states, respectively, and $\langle {S_\mathrm{BS}}^2\rangle$ and $\langle {S_\mathrm{T}}^2\rangle$ are the respective expectation values of the $\hat S^2$ operator evaluated on the KS reference states. Equation (\ref{E:projection}) works both at the strong coupling limit when $\langle {S_\mathrm{BS}}^2\rangle\sim 0$ and the BS state is essentially a singlet state as well as at the weak coupling limit when $\langle {S_\mathrm{BS}}^2\rangle\sim 1$, and should provide a reasonable estimate of exchange coupling between the two extremes.

\begin{table*}[tb]
  \caption{The exchange coupling constants ($\mathrm{cm}^{-1}$) calculated for the H--He--H model system using various H--He distances (\AA) and different approximations}
  \label{T:HHeH}
  \begin{ruledtabular}
    \begin{tabular}{cccccccccc}
      $d(\mathrm{H-He})$ &
      CASSCF &
      HF &
      HF+$G_0W_0$ &
      BLYP &
      BLYP+$G_0W_0$ &
      B3LYP &
      B3LYP+$G_0W_0$ &
      BHHLYP &
      BHHLYP+$G_0W_0$ \\
      \hline
      $1.5$ & $-1127$ & $ -909$ & $-1164$ & $-2634$ & $-1044$ & $-2150$ & $-1173$ & $-1700$ & $-1281$ \\
      $1.6$ & $ -610$ & $ -494$ & $ -624$ & $-1452$ & $ -358$ & $-1192$ & $ -542$ & $ -943$ & $ -658$ \\
      $1.7$ & $ -327$ & $ -267$ & $ -333$ & $ -800$ & $ -120$ & $ -659$ & $ -254$ & $ -521$ & $ -339$ \\
      $1.8$ & $ -175$ & $ -144$ & $ -177$ & $ -441$ & $ -368$ & $ -364$ & $ -120$ & $ -286$ & $ -176$ \\
      $1.9$ & $  -93$ & $  -77$ & $  -94$ & $ -244$ & $   -8$ & $ -201$ & $  -57$ & $ -157$ & $  -92$ \\
      $2.0$ & $  -49$ & $  -41$ & $  -50$ & $ -134$ & $    1$ & $ -110$ & $  -27$ & $  -86$ & $  -49$ \\
      $2.1$ & $  -26$ & $  -22$ & $  -26$ & $  -74$ & $    3$ & $  -61$ & $  -13$ & $  -47$ & $  -26$ \\
      $2.2$ & $  -14$ & $  -11$ & $  -14$ & $  -41$ & $    2$ & $  -33$ & $   -7$ & $  -25$ & $  -14$ \\
      $2.3$ & $   -7$ & $   -6$ & $   -7$ & $  -22$ & $    2$ & $  -18$ & $   -3$ & $  -14$ & $   -8$ \\
      $2.4$ & $   -4$ & $   -3$ & $   -4$ & $  -12$ & $    1$ & $  -10$ & $   -2$ & $   -7$ & $   -4$ \\
      $2.5$ & $   -2$ & $   -2$ & $   -2$ & $   -6$ & $    0$ & $   -5$ & $   -1$ & $   -4$ & $   -2$ \\
    \end{tabular}
  \end{ruledtabular}
\end{table*}

\section{Results and discussion}
The exchange coupling constants calculated for the H--He--H system are summarized in Table \ref{T:HHeH}. When comparing the $G_0W_0$ and DFT results to the CASSCF results, it is immediately clear that all HF and DFT results are significantly and consistently improved by inclusion of the $G_0W_0$ correction. The best results are obtained with the HF + $G_0W_0$ method.

Both with and without the $G_0W_0$ approximation, HF gives better results than the DFT methods. This can be rationalized in terms of the delocalization error present in approximate XC functionals. Delocalization error tends to overestimate the delocalization of the magnetic orbitals. This leads to overestimation of the kinetic exchange contribution and excessively antiferromagnetic coupling. Since the $G_0W_0$ approximation does not relax the orbitals, the excessive delocalization is also carried into the $G_0W_0$ results. The best DFT results are obtained with the BHHLYP functional with $50\%$ of Fock exchange and the worst with the pure GGA functional BLYP showing that reduction of exact exchange in the XC approximation deteriorates the results.

The observed trends are consistent with the observation of Phillips and Peralta\cite{peralta_2012a} that exchange coupling constants evaluated using pure GGA functionals non-self-consistently on densities obtained from hybrid calculations showed uniform improvement compared to self-consistent GGA BS-DFT calculations. Thus, similar to conventional BS-DFT calculations, the results in Table \ref{T:HHeH} clearly show that the quality of the orbitals, more specifically the presence of excessive delocalization, is critical for the quality of the $G_0W_0$ results. This further suggests that the errors in both the $G_0W_0$ and BS-DFT results are dominated by the so-called density-driven error (the self-consistent part of the calculation provides a poor estimate of the density) over the functional error (the XC approximation is incapable of providing a good energy estimate from a good density) as defined by Burke\cite{burke_2013a,burke_2017a}.

Although the results on the simple H--He--H system show consistent improvement upon inclusion of the $G_0W_0$ corrections, the situation is less clear in more complicated complexes. Table \ref{T:complexes} lists the exchange coupling constants calculated for various experimentally characterized bimetallic Cu(II) complexes. The values do not show any clear improvements in the $G_0W_0$ results as compared to the BS-DFT results. In fact, the results are more often deteriorated upon inclusion of the $G_0W_0$ correction. This discrepancy can be explained by considering the differences in the electronic structure of the simple H--He--H model and the more realistic complexes.

\begin{table*}[tb]
  \caption{The exchange coupling constants ($\mathrm{cm}^{-1}$) calculated for experimentally characterized bimetallic Cu(II) complexes using different approximations (the missing values are due to convergence issuses is the \textsc{MolGW} code)}
  \label{T:complexes}
  \begin{ruledtabular}
    \begin{tabular}{cccccccc}
      &
      Exp. &
      BLYP &
      BLYP + $G_0W_0$ &
      B3LYP &
      B3LYP + $G_0W_0$ &
      BHHLYP &
      BHHLYP + $G_0W_0$ \\
      \hline
      CUAQAC\cite{CUAQAC02a,CUAQAC02b} & $ -286$ & $ -812$ & $ -439$ & $ -361$ & $ -306$ &  $ -124$ & $-1095$  \\
      DUCGEN\cite{DUCGEN}              & $ -800$ & $-5022$ & $-5698$ & $-2164$ & $-2596$ &  $ -814$ & $  245$  \\
      EDNCOX10\cite{EDNCOX10}          & $  -75$ & $-1006$ & $-1595$ & $ -267$ & $ -779$ &  $  -53$ & $ -652$  \\
    \end{tabular}
  \end{ruledtabular}
\end{table*}

The H--He--H model includes the basic components of Anderson's superexchange model; namely, the kinetic and direct exchange contributions. The system, however, lacks necessary occupied and virtual orbitals to include spin- and dynamic charge polarization effects. It has been shown that using configuration interaction methods inclusion of such effects can make a majority contribution to the quantitative value of $J_{ij}$ in bimetallic transition metal complexes.\cite{guihery_2014a} This is not to say that the actual exchange coupling is dominated by polarization effects, but that they are important to provide a correct description of the kinetic exchange effect in the \emph{ab initio} calculations. At BS-DFT level, the kinetic exchange contribution is reasonably well described by XC functionals with a larger-than-average percentage of exact exchange\cite{peralta_2011a} which allows partial elimination of the delocalization error, which would lead to overestimation of the kinetic component to the exchange. In the H--He--H system the best results are obtained at the $100\%$ Fock exchange limit. However, in realistic systems, which also include dynamic spin- and charge-polarization effects a lower percentage of Fock exchange is necessary to include a sufficient amount of GGA exchange to correctly describe the spin- and charge-polarization effects. Calculation on the H--He--H model at $G_0W_0$ level show that the kinetic exchange contribution is correctly described as the results show a systematic improvement. Most likely, however, the spin- and charge-polarization effects outside the magnetic orbital space are not correctly described at $G_0W_0$ level, or at least they are inconsistently introduced to the BS and triplet states, leading to errors in the evaluated energy difference and to the inferior results.

\section{Conclusions}

In the present study we have applied a broken symmetry version of the $G_0W_0$ approximation to the calculation of isotropic exchange coupling constants. The results for the simple H--He--H model system show a significant and consistent improvement of the results upon inclusion of the $G_0W_0$ correction. The best results are obtained at the HF + $G_0W_0$ level showing that in the simple model system, the XC potential does not improve the results. The $G_0W_0$ correction is most useful when the orbitals are not excessively delocalized. In more realistic complexes, the $G_0W_0$  correction does not improve the DFT results. This discrepancy can be explained by poorer description of charge- and spin-polarization contributions to the exchange coupling constants at $G_0W_0$ level. Thus, at the $G_0W_0$ level of approximation, the $GW$ method only improves the results for systems where spin and charge polarization effects on the energy difference between different spin states can be neglected. In systems where the polarization effects are important, any improvement in the results is canceled by the inconsistencies in the evaluation of the polarization effects. Proper treatment of the polarization contribution to the energy difference most likely requires self-consistent calculations. It should be noted that eigenvalue-self-consistent calculations where attempted but they did not lead to visible improvement of the results.

The results nonetheless shows the clear potential of $GW$ approaches for improving the description of exchange coupling constants at reasonable computational costs. It is well-known that the evaluation of total energies (which are only involved for the calculation of $J$ by BS approaches) improves systematically when the quasiparticle wave functions are calculated self-consistently in the $GW$ potential ($GW_0$ approximation) and further with self-consistent evaluation of the polarization function and screened electronic repulsion.\cite{martin_2016} Furthermore, self-consistency removes the dependence of the results on any approximate XC functional. The effect of full self-consistency on the evaluation of $J$ in BS $GW$ calculations will be addressed in a subsequent publication.

\section*{Acknowledgement}
Funding for this work has been provided by the Magnus Ehrnrooth Foundation (A.\,M.), Kvantum Institute of the University of Oulu (A.\,M.), the China Scholarship Council (Z.\,H) and the scientific research grant R-143-000-A80-114 of the National University of Singapore (N.\,I.).
The computational resources and services were provided by the CSC-IT Center for Science in Finland and the Finnish Grid and Cloud Infrastructure (persistent identifier urn:nbn:fi:research-infras-2016072533).

\section*{Data availability}

The data that support the findings of this study are available from the corresponding author upon reasonable request.

\bibliography{references}

\end{document}